\documentclass{cimento}

\newcommand{\powheg}{{\sc POWHEG}}
\newcommand{\alpgen}{{\sc ALP\-GEN}}

\newcommand{\ttbar}{\ensuremath{t{\overline{t}}}}
\newcommand{\bbbar}{\ensuremath{b{\overline{b}}}}
\newcommand{\tbar}{\ensuremath{\overline{t}}}
\newcommand{\pt}{\ensuremath{p_{T}}}
\newcommand{\roots}{$\sqrt{s}$}
\newcommand{\mttbar}{\ensuremath{m_{t\bar{t}}}}
\newcommand{\mtw}{\ensuremath{m_{\mathrm{T}}^{W}}}
\newcommand{\invfb}{$\mathrm{fb}^{-1}$}
\newcommand{\intlum}{$\int${\it L}dt}
\newcommand{\lnu}{\ensuremath{\ell\nu_\ell}}
\newcommand{\epem}{\ensuremath{e^+e^-}}
\newcommand{\mupmu}{\ensuremath{\mu^+\mu^-}}
\newcommand{\sigmattbar}{$\sigma_{\ttbar}$}
\newcommand{\wjets}{\ensuremath{W\!+\!\mathrm{jets}}}
\newcommand{\zjets}{\ensuremath{Z\!+\!\mathrm{jets}}}
\newcommand{\muplus} {\ensuremath{\mu+\mathrm{jets}}}
\newcommand{\eplus} {\ensuremath{e+\mathrm{jets}}}

\def\ipb{\mbox{pb$^{-1}$}}
\def\TeV{\ifmmode {\mathrm{\ Te\kern -0.1em V}}\else
                   \textrm{Te\kern -0.1em V}\fi}%
\def\ETmiss{\ensuremath{E_{\mathrm{T}}^{miss}}} 
\def\Zboson{\ensuremath{Z}}                                                                        
\def\Wboson{\ensuremath{W}}%

\usepackage{wrapfig}
\usepackage{graphicx}
\usepackage{amsmath}
\usepackage{amssymb}
\usepackage{parskip}
\usepackage{mathtools}  
\mathtoolsset{showonlyrefs}  

\usepackage{lineno}

\title{Studies of top quark production with the ATLAS detector at the LHC}

\author{F.~Span\`o, On behalf of the ATLAS Collaboration}
\instlist{\inst{} Royal Holloway, University of London, UK}

\PACSes{\PACSit{14.65}{ha}}

\begin{document}

\maketitle
\vspace{-0.9cm}
\begin{abstract}
A review is presented of the most recent measurements of top quark strong and electroweak
production performed by using data collected with the ATLAS detector
in proton-proton collisions at \roots\ = 7 and 8 TeV at the Large Hadron Collider
corresponding to integrated luminosities of up to about 4.7 \invfb\ and 20 \invfb, respectively.
\end{abstract}

\vspace{-0.8cm}
\section{Why top quark?}
\vspace{-0.2cm}
The top quark ($t$) is the most massive known elementary particle. Its
large expected Yukawa coupling with the newly discovered Higgs boson
gives it a prominent role in the electroweak symmetry breaking
mechanism. The measurement of its production and decay at colliders
is then a crucial test of the standard model (SM) and a probe
for new physics. On one hand understanding top quark production
allows to control a crucial background both to Higgs boson production and to
possible new physics like supersymmetry.
 On the other hand, many new physics scenarios, from the presence of extra
dimensions to the existence of new strong forces, require direct or
indirect couplings of new particles to top quarks that could modify the SM
top quark production cross section.
\vspace{-0.4cm}
\section{Top producer and observer}
\vspace{-0.4cm}
Abundant production of top quarks is desirable for a detailed study of their properties. 
The Large Hadron Collider~\cite{ref:lhc} is a prolific top quark
producer.  During its first run period (Run 1), the LHC collided
protons at the unprecedented center-of-mass
energies (\roots) of 7 and 8 TeV, respectively in 2010-2011 and in
2012. Run 1 was characterized by smooth running conditions and a
vertiginous increase in peak luminosity from  $2.1\cdot
10^{32}\,\mathrm{cm}^{-2}\,\mathrm{s}^{-1}$ in 2010 to the
peak value of  $7.7\cdot10^{33}\,\mathrm{cm}^{-2}\,\mathrm{s}^{-1}$ in
2012. The integrated luminosity
(\intlum) delivered per collision point was about 5 \invfb\ in 2011 and 22
\invfb\ in 2012,  respectively 100 and 440 times larger than the one
delivered in 2010 (50 \ipb).
\newline 
\noindent The ATLAS detector~\cite{ref:atlas} is an excellent top observer.
A multi-purpose composite detector with cylindrical symmetry around the
line of collision of the proton beams, it covers nearly the
entire solid angle surrounding the collision point.
From the innermost layer outwards, it features an inner tracking
detector, surrounded by a thin superconducting solenoid magnet producing a 2 T
axial magnetic field, followed by electromagnetic and hadronic
calorimeters and by an external muon
spectrometer incorporating three large toroid magnet assemblies.
The full detector is at play in reconstructing events with top quarks, efficiently
selected by a three-level trigger system which reduces the overall event rate to
about 400 Hz into storage.
\vspace{-0.5cm}
\section{Top quark production and decay at LHC: predictions} 
\vspace{-0.4cm}
In LHC proton-proton collisions top quarks are predominantly strongly
produced in top-antitop (\ttbar) pairs. At \roots\ = 7 (8) TeV the largely enhanced gluon-gluon fusion
process dominates over the quark-quark annihilation process (85\% vs 15\% of the total
\ttbar\ cross section at leading order) and it boosts the \ttbar\ production cross
section, \sigmattbar, to a value that is 25 (35) times larger than
that foreseen and observed in proton-antiproton collisions at \roots\ = 1.96 TeV
at the Tevatron. The predicted value of \sigmattbar\ is now known at the level of
about 4\% thanks to recent breakthrough calculations including all
Quantum Chromodynamics (QCD) effects at NNLO, complemented with NNLL resummation~\cite{Czakon:2013goa}.
At a rate of about half the \ttbar\
production~\cite{ref:tchan,ref:Wtchan,ref:schan}, the top quark is
also produced singly; such production is described (at leading order) by three
electroweak processes. The \Wboson\ boson mediated  $t$- and $s$-channel account respectively for
75\% and 5\% of the total single top quark production cross section,
$\sigma_{t,total}$, while the remaining 20\% is accounted for by the
associated production of a single top quark and a \Wboson\ boson, mediated
by a $b$-quark.
Thanks to the large top quark production cross sections, the LHC
produced about 5.4 (0.96) million \ttbar\ events and 2.5 (0.47)
million single top quark events at \roots\ = 8 (7) TeV. 
\newline
\noindent As the top quark decays to a \Wboson\ boson and a $b$-quark about 100\% of the times, 
 the final state of a \ttbar\ event is characterized by the number of
 \Wboson\ bosons decaying
to a lepton-neutrino pair~\footnote{The \Wboson $\rightarrow$\lnu\
  ($q\overline{q'}$) decay takes place 32.4\%  (67.6\%) of the times where
  $q$ is a light quark, $\ell$ is a lepton, $\nu_{\ell}$ is the
  corresponding lepton neutrino.}. 
The case when both \Wboson\ bosons decay hadronically represents 45.7\% of
the events (fully hadronic channel), while 44.1\% of the events feature
only one \Wboson\ boson decaying to a lepton ($\ell$) and the corresponding neutrino
($\nu_{\ell}$) ( 34.3\%  of the times $\ell$ is an electron ($e$) or a muon
($\mu$),  or a tau ($\tau$) decaying to an $e$ or $\mu$ ) (single lepton
channels).  The leptonic decay of both \Wboson\
bosons to $e$, $\mu$  or $\tau \rightarrow$ $e/\mu$ takes place 6.5\% of the
time while the remaining 3.7\% corresponds to hadronic
$\tau$ decays associated with a lepton.
The \ttbar\ final state consists of $b$-jets from $b$-quark production, high \pt\
jets from hadronic \Wboson\ boson decays,  at least one or two high $p_{t}$
leptons and large missing transverse energy
(\ETmiss) due to the neutrino in \Wboson\ boson leptonic decays.
The \ttbar\ and single top quark final states have similar backgrounds (single
bosons (\Wboson,~\Zboson) plus jets, dibosons and multi-jet events)
and they are background to each other.
\vspace{-0.4cm}
\section{Top quark pair production}
\vspace{-0.4cm}
\subsection{Inclusive $\sigma_{\ttbar}$  at \roots\ = 8 TeV}
\label{sec:dilepXsec8TeV}
The latest ATLAS measurement of \sigmattbar~\cite{ref:dilep} uses
dilepton events selected from the full dataset collected  at \roots\
= 8 TeV (\intlum\ = 20.3~\invfb) by requiring a pair of central, high momentum, opposite-sign  $e$ and $\mu$. 
The dominant single top quark background in the $Wt$-channel is
derived from simulation. The estimates of the smaller fake lepton and
\zjets\ backgrounds are data-driven by extrapolating the content of
background-enriched control regions in same sign lepton pair events
and in \Zboson\ $\rightarrow$ \mupmu\ events respectively.
The contents of the two highly \ttbar-pure samples obtained 
\setlength{\intextsep}{4pt}
\setlength{\columnsep}{6pt}
 \begin{wrapfigure}[17]{l}{0.45\textwidth}
\centerline{
   \includegraphics[width=0.38\textwidth]{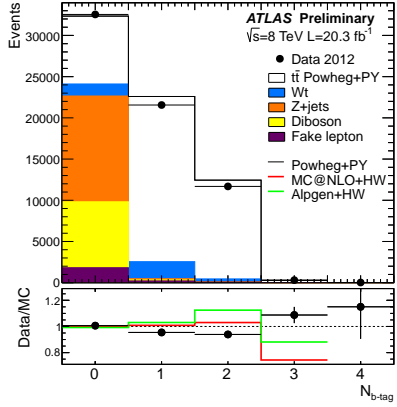}   
 }
\caption{Distribution of the number of $b$-tagged jets in preselected
  opposite-sign $e$-$\mu$ events from ref.~\cite{ref:dilep}. Data
  is compared to predictions from different MC generator codes
  for \ttbar\ events.}
\label{fig:dilepXsec}
 \end{wrapfigure}
by requiring exactly one and two $b$-tagged jets (shown in
fig.~\ref{fig:dilepXsec}) are expressed as functions of $\sigma_{\ttbar}$ and the efficiency to
select, reconstruct and $b$-tag a jet, $\epsilon_{btag}$. A simultaneous
fit to the content of the two samples is performed to extract $\sigma_{\ttbar}$ and
$\epsilon_{btag}$  and to use the data information to constrain
jet-related and $b$-tagging systematic uncertainties.
The resulting \ttbar\ cross section is 
$\sigma_{\ttbar} = 237.7 \pm 1.7\,\mathrm{(stat.)}\pm 7.4\,\mathrm{(syst.)} \pm 7.4\,\mathrm{(lumi)} \pm 4.0\,\mathrm{(beam\,  energy)\, pb}$
in agreement with the SM prediction. 
The relative total uncertainty of 4.8\%  is dominated by systematic
effects originating mostly from the limited knowledge  of
the total integrated luminosity and of the beam energy, with the other
two major contributions coming from \ttbar\ modelling and electron
isolation efficiency.
\newline
\newline
\begin{figure}[htbp]
\centerline{
\includegraphics[width=0.44\textwidth]{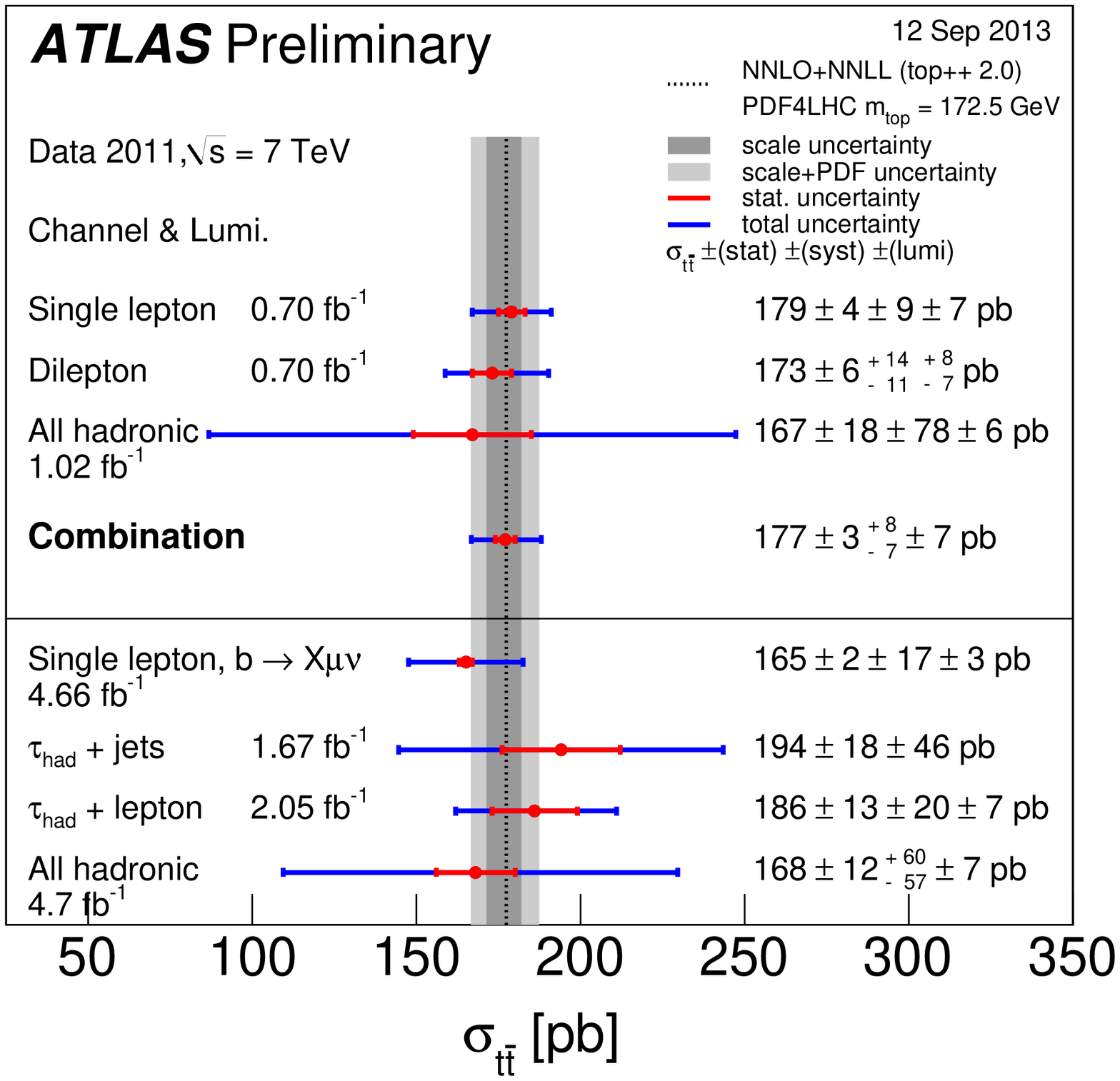}    
\includegraphics[width=0.44\textwidth]{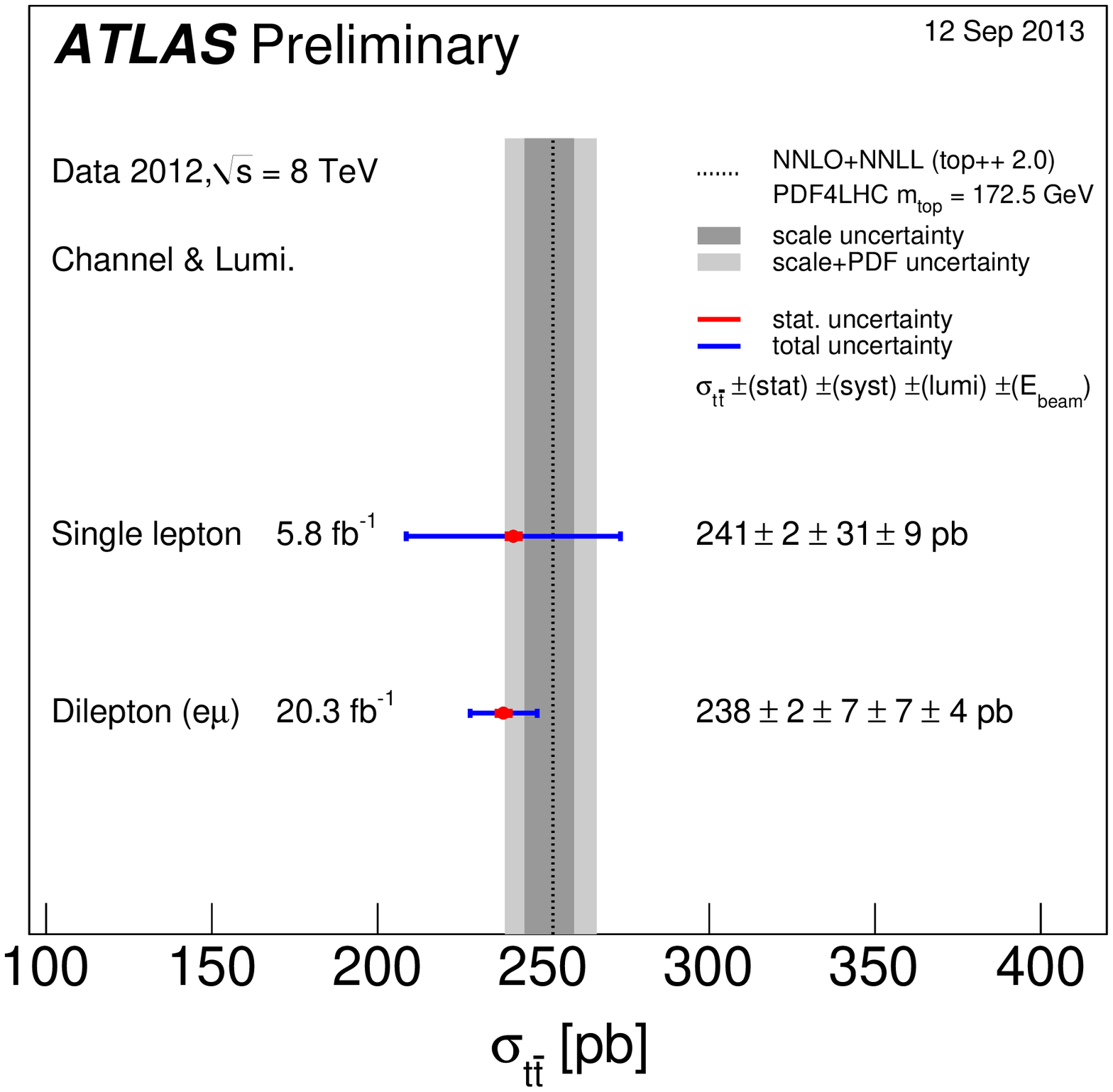}    
}
\caption{Summary of ATLAS measurements of \sigmattbar\ for \roots = 7 TeV (left) and \roots = 8 TeV (right)
compared to the NNLO+NNLL QCD calculation~\cite{ref:xsecCom7TeV,ref:xsecCom8TeV}. On the left, the upper part of the
figure shows measurements that contribute to the combined value~\cite{ref:ttbarXsecCom7TeV}
 while the lower part shows additional newer measurements which are not included in the combination.}
\label{fig:ttbarXsecSummary}
\end{figure}
\vspace{-0.8cm}
\noindent
\subsection{Summary of inclusive $\sigma_{\ttbar}$  at \roots\ = 7 and
  8 TeV}
The very precise dilepton measurement of \sigmattbar\ outlined in
sect.~\ref{sec:dilepXsec8TeV} is one of the many results obtained using a variety of final state samples at \roots\ =
7 and 8 TeV~\cite{ref:xsecCom7TeV,ref:xsecCom8TeV} and illustrated in
fig.~\ref{fig:ttbarXsecSummary}.
The measurements are dominated by systematic uncertainties and they
are well consistent with the latest NNLO+NNLL QCD
predictions~\cite{Czakon:2013goa}. 
The relative uncertainty, $\delta$\sigmattbar/\sigmattbar, of the most precise results,
the ATLAS combination at \roots\ = 7 TeV
($\delta$\sigmattbar/\sigmattbar\ = 6.2\%) and the dilepton
measurement at \roots\ = 8 TeV ($\delta$\sigmattbar/\sigmattbar\ = 4.8\%) is comparable with the theoretical relative uncertainty of about 4\%.
ATLAS also contributes to the first LHC \sigmattbar\ combination~\cite{ref:ttbarXsecCom7TeV} at
\roots\ = 7 TeV which gives
$\sigma_{\ttbar} = 173.3 \pm 2.3\,\mathrm(stat.) \pm 7.6\,\mathrm{(syst.)} \pm 6.3\,\mathrm{(lumi)\,pb}$
with a relative total uncertainty of 5.8\%.
\vspace{-0.4cm}
\subsection { $\sigma_{\ttbar + heavy flavour}$  at \roots\ = 7 TeV}
\begin{figure}
\centerline{
\includegraphics[width=0.42\textwidth]{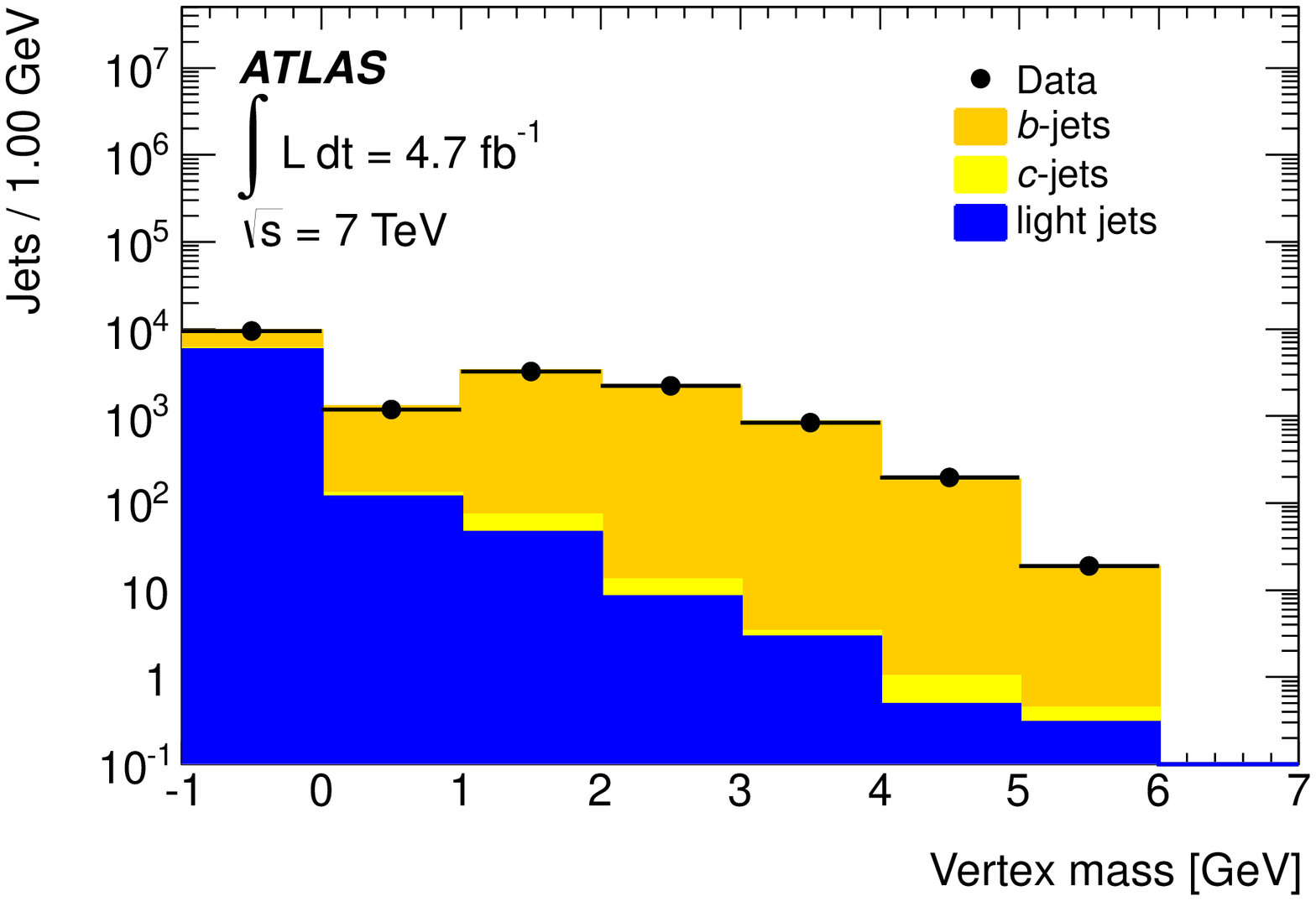}    
\includegraphics[width=0.42\textwidth]{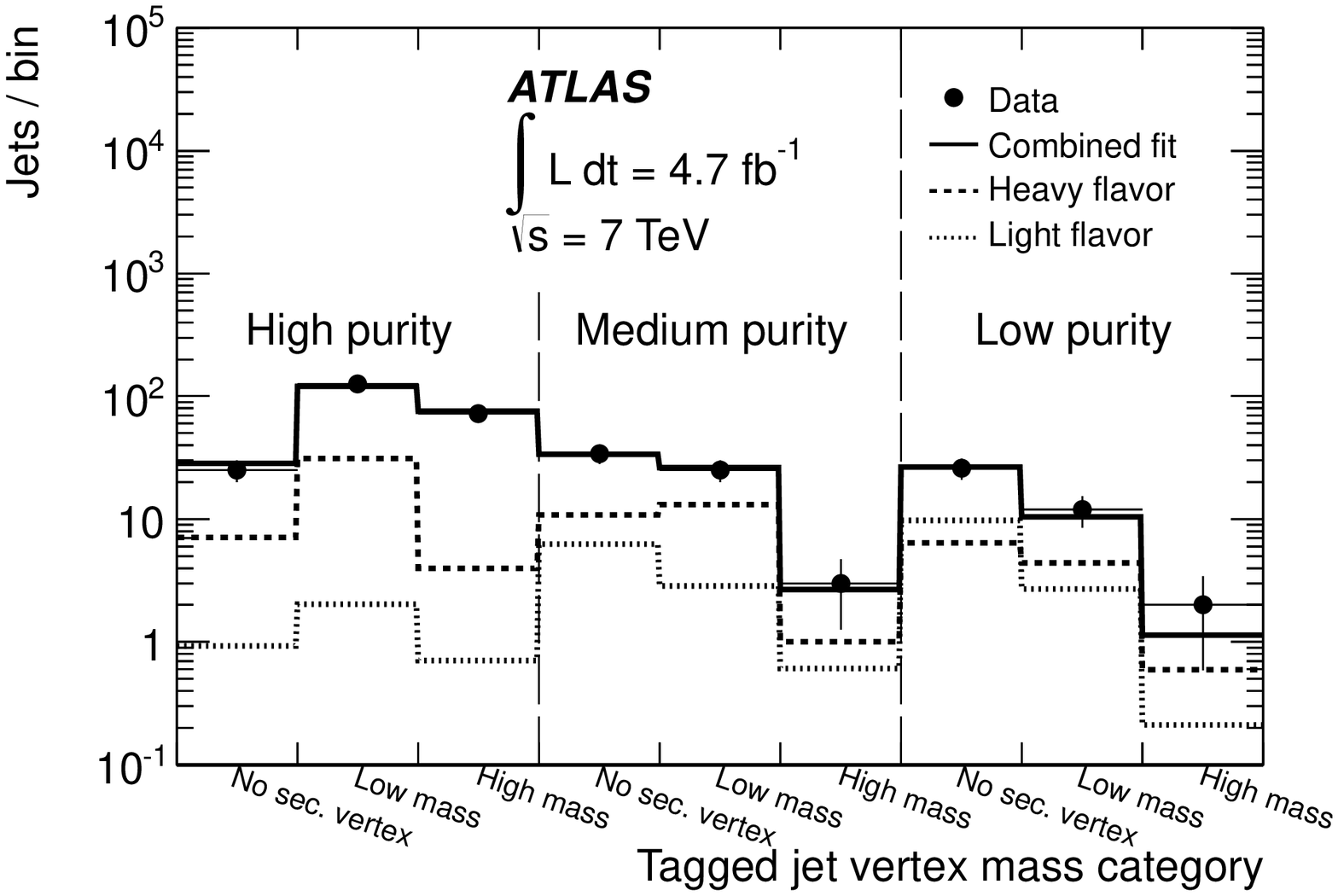}    
}
\caption{
Left: vertex mass distribution for high purity $b$-tagged
  jets~\cite{ref:ttPlusHf} for dilepton selected events with no
  requirement on $b$-tagged jet multiplicity. Data is compared to predictions.
  Jets with no reconstructed secondary decay vertex are
  assigned to the `-1 GeV' bin. Right: result of the template fit
  (solid line) to the vertex mass distribution in data (points)~\cite{ref:ttPlusHf}. Data
  is divided into three groups depending on the purity of $b$-jets
  passing each selection, as described in ref.~\cite{ref:ttPlusHf}. 
The best fit is shown as the sum
  (labeled as `Combined fit', which includes the $b$-jets from top quark
  decay) of separate contributions from additional $b$- and $c$-jets
  (labeled as `Heavy flavor'), and LF (labeled as `Light
  flavor').  Within each purity category, the first bin contains jets
  with no secondary vertex; the middle (third) bin contains jets
  with `low' ('high') mass i.e. smaller (larger) than 2 GeV. 
 }
\label{fig:ttPlusHFfigVMass}
\end{figure}

Going to a more exclusive final state, the cross section for the production
of a top quark pair in association with at least one $b$- or $c$- quark is
extracted using the full dataset collected at~\roots\ = 7
TeV (\intlum\ = 4.7~\invfb)~\cite{ref:ttPlusHf}. The \ttbar\ + $b$ or $c$ + X final state is the main background to the associate production
of a top quark pair with a Higgs boson decaying to
a \bbbar\ quark pair. Events are selected in the dilepton channel by
requiring two opposite-sign leptons ($e$, $\mu$) and at least two central high $p_{T}$
jets. Appropriate requirements on large \Wboson\ boson transverse mass
(\mtw), dilepton mass and $H_{T}$, the sum of the $p_{T}$ of all jets and leptons in the event,
are used  to select same flavour ($\epem$, $\mupmu$) and opposite flavour
($e^{\pm}\mu^{\mp}$) leptons.  The backgrounds are estimated and subtracted in a
similar way as those for the dilepton measurement outlined in sect.~\ref{sec:dilepXsec8TeV}.
Two main ingredients are then determined.
In the subset of events with at least three $b$-tagged jets the cross
section for the production of a \ttbar\ pair in association
with heavy flavours (at least one $b$- or $c$-quark) is calculated as 
$\sigma_{fid} (tt+HF)$= $\frac{N_{HF}}{\int L dt \cdot
  \epsilon_{HF}}$ in fiducial form $i.e.$ such that the
acceptance correction $\epsilon_{HF}$  is calculated with respect to a phase
space volume close to the experimentally selected one.
The number of $b$-tagged jets arising from additional
heavy flavours with respect to the \ttbar\ system, $N_{HF}$, is obtained by a
maximum likelihood fit of templates of jets with different origin
(\ttbar\, additional heavy flavour, non-\ttbar\ and mis-tagged jets)
to the two dimensional distribution of $b$-tagged jet \pt\ and displaced
vertex mass. The distribution of the latter variable, displayed in fig.~\ref{fig:ttPlusHFfigVMass},  is
shown to be sensitive to $b$- , $c$-jet and light jet content.
 The fit is carried out using three samples with increasing
 $b$-tagging purity to enhance the separation between light and heavy
 flavour jets (as shown in fig.~\ref{fig:ttPlusHFfigVMass} on the right).
The selected 106 events feature about 325 jets whose large uncertainty
on the $b$- and $c$-jet separation content only allows to measure the combined
heavy flavour content. The $N_{HF}$ value is then converted into 
$\sigma_{fid} (tt+HF)$ by scaling it with $\int L dt$ and the
simulation-derived efficiency $\epsilon_{HF}$ for selecting \ttbar\ events with at
least three $b$- or $c$-matched jets (two of which come from the \ttbar\ pair).
The second ingredient is the fiducial cross section for the
production of a top quark pair associated with at least one additional jet,
$\sigma_{fid} (tt+j)$= $\frac{N_{j}}{\int L dt \cdot \epsilon_{j}}$.
The value of $\sigma_{fid} (tt+j)$  is determined by counting $N_{j}$,
the number of events with at least three jets, at least two
of which are $b$-tagged, and by scaling it with $\int L dt$ and the simulation-derived efficiency $\epsilon_{j}$ for
selecting \ttbar\ events with at least three jets (two of which
come from the \ttbar\ pair).
The two ingredients are then combined to produce the cross section
ratio $R_{HF} = \frac{\sigma_{fid} (tt+HF)}{\sigma_{fid} (tt+j)} =
6.2\pm1.1\,\mathrm{(stat.)}\pm 1.8\,\mathrm{(syst.)} \% $
which is consistent with SM predictions ranging from 3.4\%
(\alpgen{}) to 5.2\% (\powheg).
The relative uncertainty is dominated by systematic uncertainties on $c$-tagging
efficiency ($\approx $ 21\%), hadronization ($\approx$ 10\%) and flavour
composition($\approx$ 6\%).
\vspace{-0.4cm}
\subsection{Differential  \sigmattbar\ in lepton plus jets events at  \roots\ = 7 TeV}
The large number of
\ttbar\ events allows to measure differential \ttbar\ cross sections in the
lepton plus jets sample using the full dataset collected at \roots\  = 7
TeV~\cite{ref:diffxsec7TeV}. 
Events are selected by requiring at least one isolated
high $p_{T}$ lepton ($e$ or $\mu$),  large \mtw\ and \ETmiss,
consistent with only one leptonically decaying
\Wboson\ boson, plus at least four high \pt\ central jets and at
least one $b$-tagged jet, due to the full hadronic decay of one top quark and
the one $b$-jet  in the decay of the other top quark. 
After estimating the dominant \wjets\ background with data-driven
techniques (using the charge-asymmetric \Wboson\ boson production at LHC) and the
remaining small background with a mix of simulation (for single top quark and
dibosons) and data-driven techniques (for events with fake leptons),
the full \ttbar\ topology is reconstructed  with a kinematic
likelihood fit incorporating the knowledge of the \Wboson\ boson
mass and the constraint of same mass for top and antitop quarks.
\begin{figure}
\centerline{
\includegraphics[width=0.40\textwidth]{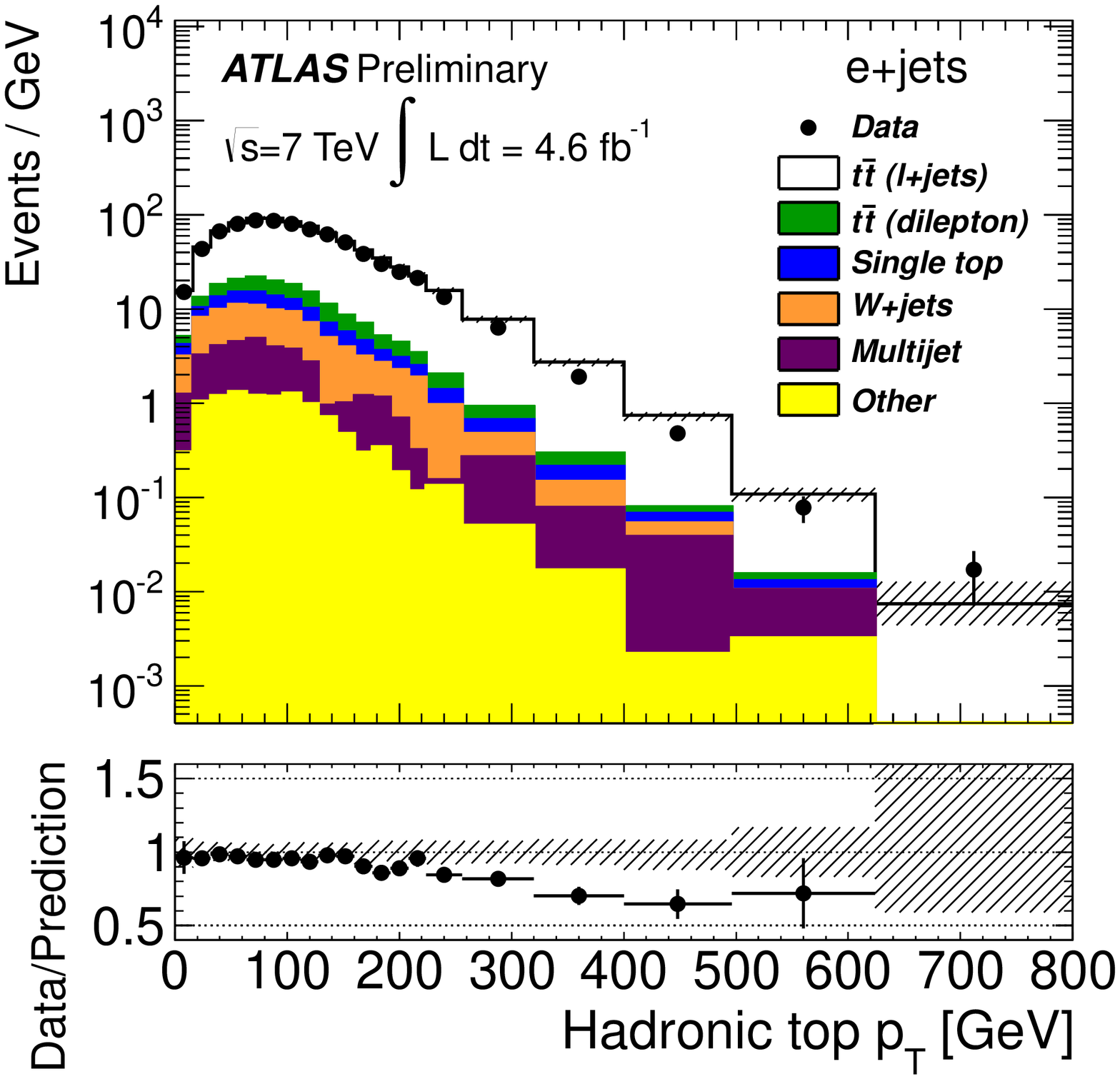}    
\includegraphics[width=0.47\textwidth]{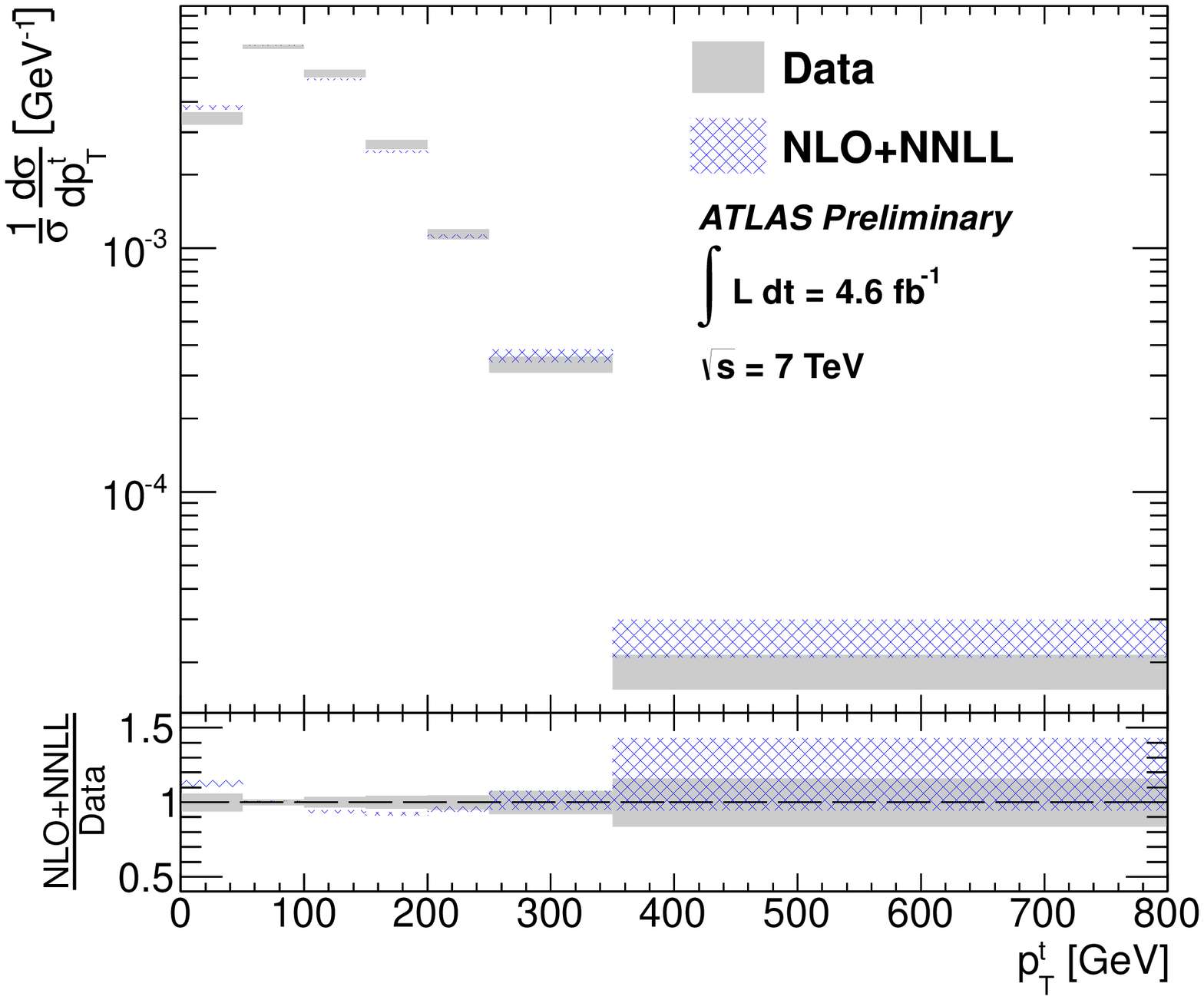}    
}
\caption{Left: distribution for reconstructed $p_{T,top}$ (from the
  hadronically decaying top quark) in the $e$+jets
  channel~\cite{ref:diffxsec7TeV}. Data is
  compared to predictions. The hashed area represents the combined
  statistical and  systematic uncertainties on the prediction,
  excluding systematic uncertainties related to \ttbar\ modeling.
  "Other" includes small backgrounds from diboson and
  \zjets\ events. Events beyond the axis range are included in the
  last bin. Right: normalized differential \ttbar\ cross-section as a
  function of $p_{T,top}$~\cite{ref:diffxsec7TeV} compared to NLO+NNLL
  prediction. The blue band corresponds to the fixed scale
  uncertainties on the theoretical prediction. The gray band indicates
  the total uncertainty  on the  data in each bin. }
\label{fig:topPtDiffXsec7TeV}
\end{figure}
After background subtraction the distribution of the number of events
is determined as a function of kinematic variables:  the \pt\ of the individual top quark,
$p_{\mathrm{T},top}$, the mass (\mttbar),
the rapidity ($y_{\ttbar}$) and transverse momentum
($p_{\mathrm{T},\ttbar}$) of the \ttbar\ system.
The distributions are corrected for detector and acceptance effects (unfolded) to the full
parton phase space, normalized to their integral (the total measured
cross section) and the the corresponding normalized differential cross
section are extracted.
An example for $p_{\mathrm{T},top}$  is shown in
fig.~\ref{fig:topPtDiffXsec7TeV} illustrating the data-prediction comparison
for the reconstructed distribution and the comparison of the unfolded result to the
approximate NNLO prediction. The final normalized differential cross sections are obtained by
 combining the results in the \eplus\ and in the \muplus\ samples 
\begin{figure}
\centerline{
\includegraphics[width=0.48\textwidth]{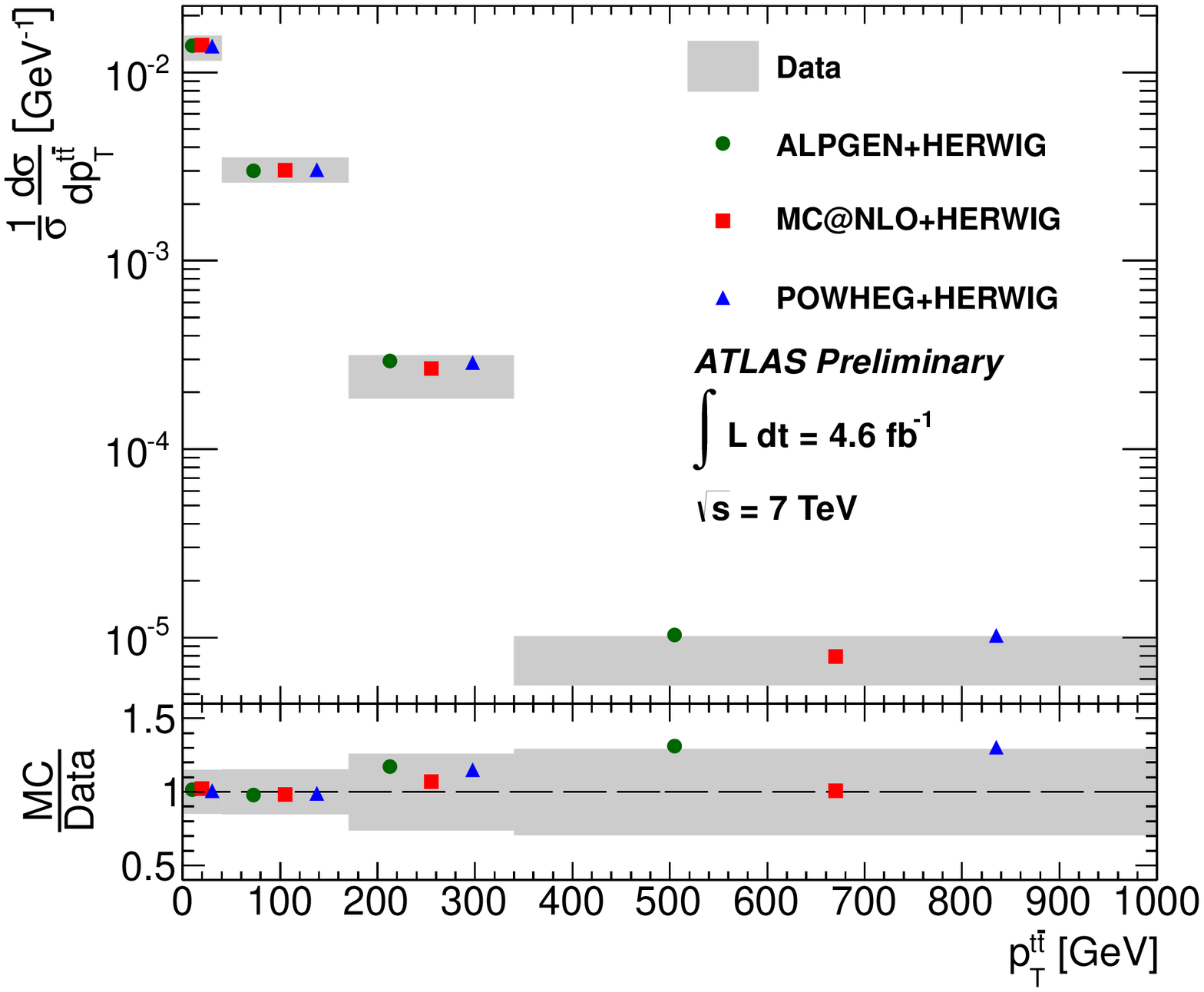}     
\includegraphics[width=0.43\textwidth]{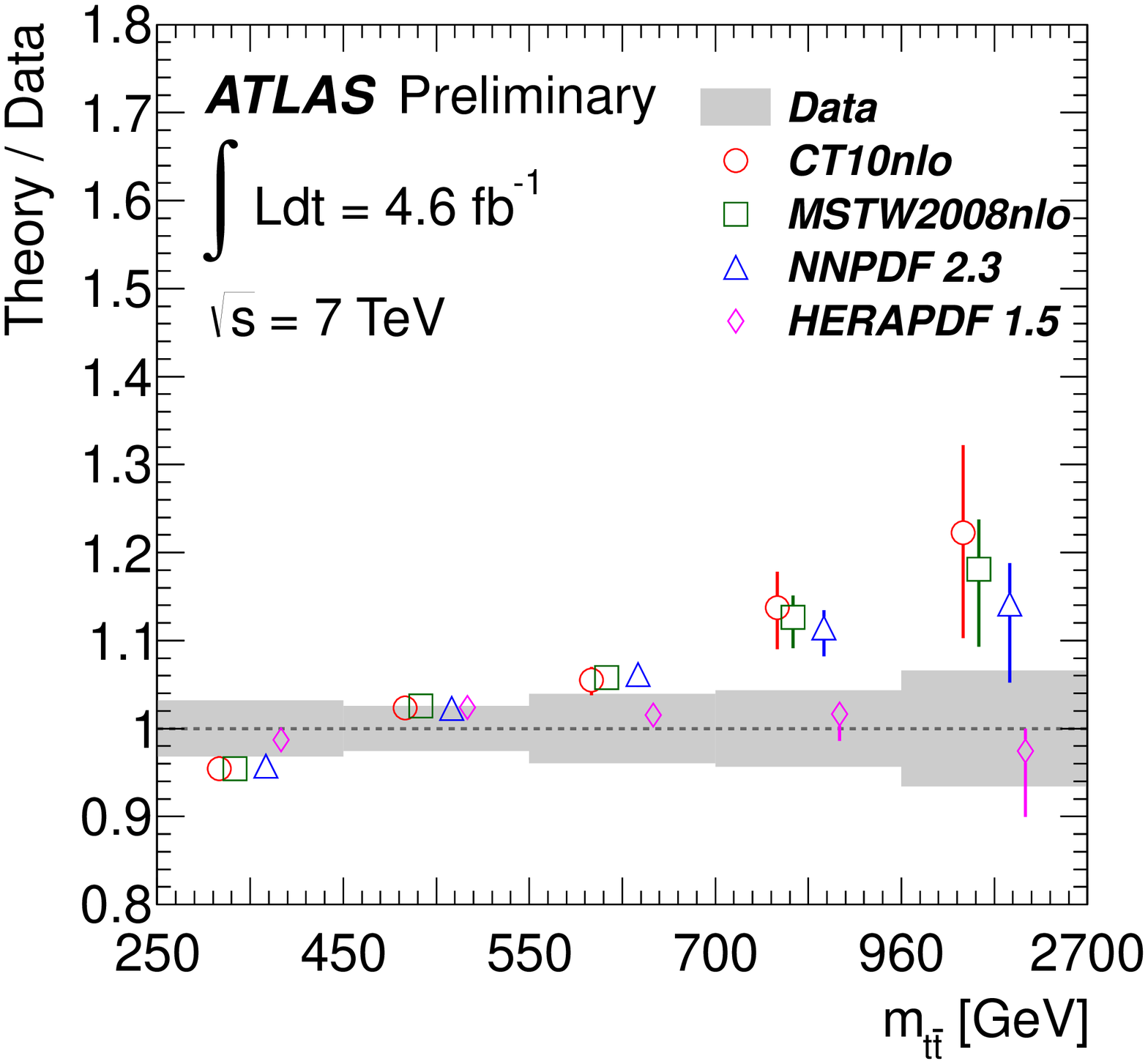}     
}
\caption{Left: normalized differential cross-sections as a function of
  $p_{\mathrm{T},top}$ compared to predictions from several generators. 
Right: ratio of NLO QCD predictions with varying PDFs to normalized
differential cross-sections as a function of \mttbar. The error bars
denote the uncertainties on the PDFs. In each bin of  both
figures~\cite{ref:diffxsec7TeV} the points are slightly offset  to
allow for better visibility and the gray band indicates the total
uncertainty on the data.}\label{fig:topDiffxsecPtandPDF}
\end{figure}
using a minimum covariance estimator including correlations.
 Systematic uncertainties are propagated through the
 unfolding corrections by modifying the migration matrix and acceptances
while keeping the data fixed in the measurement.
The results are dominated by systematic uncertainties varying from
2\% to 20\% depending on the bin and the observable.
The highlights of a large number of comparisons with predictions produced
with NLO and LO generators matched to parton showers and
NLO and approximate NNLO calculations are shown in
fig.~\ref{fig:topDiffxsecPtandPDF}. 
The analyses show potential
sensitivity to differences in predictions and in choices
of parton distribution functions (PDFs): the figure on the left
indicates a qualitative tendency to measure smaller \ttbar\ cross
sections at higher $p_{\mathrm{T},top}$ values compared to most
predictions, while the figure on the right shows a qualitative preference of
the data for the HERA-based PDF.
\vspace{-0.4cm}
\section{Single top quark production cross section}
\vspace{-0.4cm}
\subsection{Inclusive single top t-channel cross section at \roots\  =
  7 and 8 TeV}
\label{sect:singleTop7TeV}
Single top quark production was initially observed in the $t$-channel
with an integrated luminosity of 1.04 \invfb
collected at \roots\ = 7 TeV~\cite{ref:tchanPap7TeV} by requiring
events with one isolated
high $p_{T}$ central lepton ($e$ or $\mu$), exactly two or three
jets within a large pseudorapidity range ($|\eta|<4.5$) resulting
from the $b$-quark, and the accompanying jet(s) and large \ETmiss\ and
\mtw\ from the leptonic \Wboson\ boson decay.
The dominant \ttbar\ and \wjets\ backgrounds are obtained from
simulation while the fake lepton contribution is derived by
data-driven techniques.
The value of the t-channel single top quark production cross section,
$\sigma_{t,inc}$, is obtained using a binned maximum likelihood fit to the distributions of the output of a neural network
(NN) encapsulating the properties of 12 and 18 kinematic variables in
two- and three-jet event samples, respectively.
The resulting 
$\sigma_{t,inc}=  83 \pm 4\,\mathrm{(stat.)}^{+20}_{-19}\,\mathrm{(syst.)\,pb} $
is consistent with the SM prediction and its uncertainty (24\%
relative precision) is dominated by systematic effects, particularly
by initial and final state radiation ($\approx$ 9 \%) and jet energy
scale ($\approx$  7\%).
The value of the CKM matrix element $|V_{tb}|$ relating the measured and
the predicted  SM $t$-channel cross sections is found to be $V_{tb} = 1.13^{+0.14}_{-0.13} $.
\setlength{\intextsep}{-14pt}
\begin{figure}[ht]
 \includegraphics[width=0.43\textwidth]{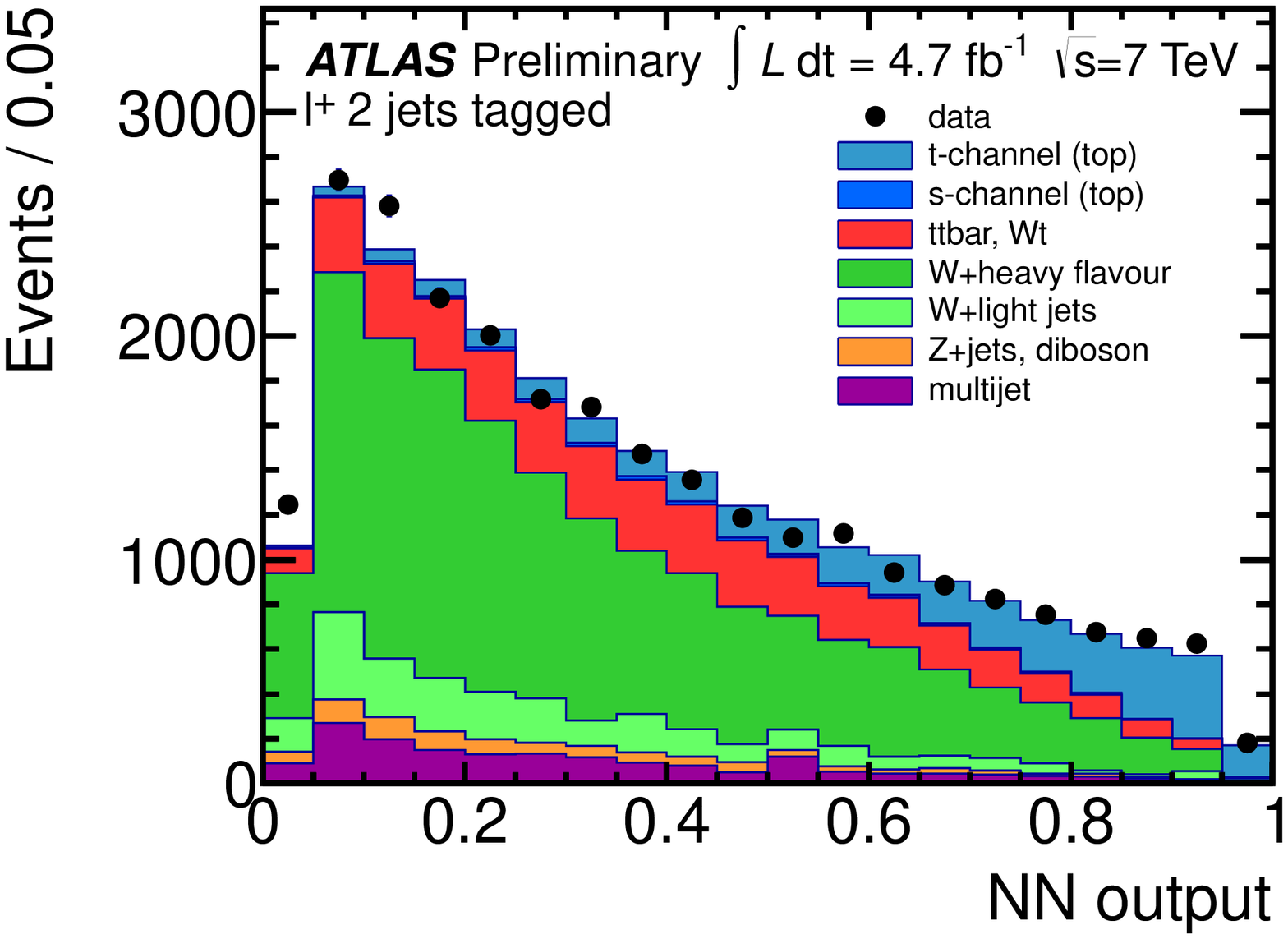}   
 \includegraphics[width=0.48\textwidth]{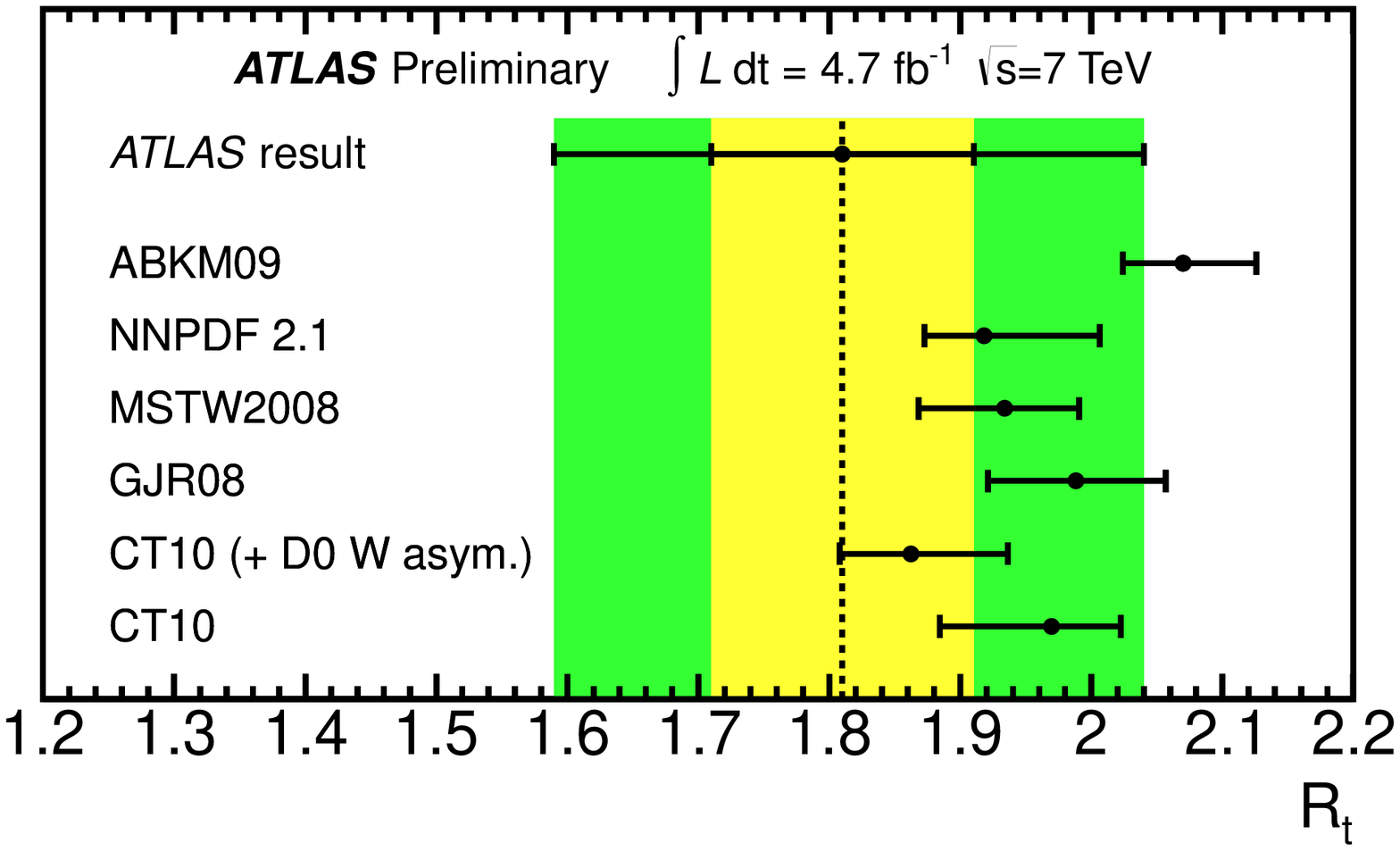}    
 \caption{Left: neural network output distribution normalized to the result
   of the binned maximum-likelihood fit in the 2-jet tagged positive
   lepton dataset. Right: comparison of measured
   $\mathrm{R}_{\mathrm{t}}$  value (black line)
   with predictions for different NLO PDF sets. The error bar represents the uncertainty on the renormalisation and factorisation scales. The statistical (combined statistical and systematic) uncertainty of the measurement is shown in yellow (green). Both
   figures derive from the t-channel measurements at \roots\ = 7 TeV from ref.~\cite{ref:tchan7TeV}.}
 \label{fig:tchanXsec7TeV}
 \end{figure}
\newline
Using the full dataset collected by ATLAS at \roots\ = 7 TeV,  the
separate production cross sections for single top $\sigma_{t}$ and single
antitop $\sigma_{\tbar}$ are obtained~\cite{ref:tchan7TeV} by using
the fit to the same NN output distribution, but for samples containing negative and
positive leptons, respectively. An example of  data-prediction 
comparison for the NN ouput distribution is shown in
fig.~\ref{fig:tchanXsec7TeV} (left) for events with positive leptons.
The results are 
$\sigma_{t} =  53.2 \pm 1.7\,\mathrm{(stat.)} \pm
10.6\,\mathrm{(syst.)\,pb}$  and $\sigma_{\tbar} =  29.5 \pm 1.5\,\mathrm{(stat.)} \pm 7.3\,\mathrm{(syst.)\,pb}$.
 They are consistent with the SM predictions and 
have a relative precision of  20\% and 25\% respectively, dominated again by jet-related systematic uncertainties. The cross section ratio $\mathrm{R}_{\mathrm{t}}$ = $\sigma_{t}$/$\sigma_{\tbar}$
\setlength{\intextsep}{-4pt}
\setlength{\columnsep}{6pt}
 \begin{wrapfigure}[20]{l}{0.50\textwidth}
\includegraphics[width=0.40\textwidth]{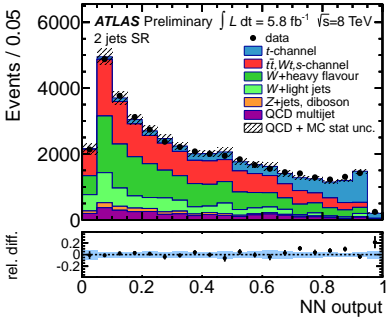}    
\caption{Neural network output distribution for the two-jet
  sample for $t$-channel observation at \roots\ = 8
  TeV~\cite{ref:tchan8TeV}. Data is compared to signal and
  backgrounds, normalized to the fit result. The lower panel shows the 
  the relative difference between observed data and prediction.
 The blue-shaded band reflects the uncertainty from the limited amount
 of simulated events and the uncertainty on the QCD multijet normalization.}
\label{fig:tchanXsec8TeV}
\end{wrapfigure}
(sensitive to the ratio of the $u$-quark to $d$-quark PDFs) is also fitted
by including correlation effects to provide a reduced uncertainty
value $\mathrm{R}_{\mathrm{t}} = 1.81 \pm
0.10\,\mathrm{(stat.)}^{+0.21}_{-0.20}\,\mathrm{(syst.)}$
The uncertainty on $\mathrm{R}_{\mathrm{t}}$ is dominated by systematic
effects (from higher order radiation modelling and experimental
jet-related effects) as one can observe in
fig.~\ref{fig:tchanXsec7TeV} (right):
the measured $\mathrm{R}_{\mathrm{t}}$ value is compared to predictions using
different NLO PDFs and its statistical
uncertainty, the spread of the predictions and their
uncertainties are of the same order of magnitude.
\noindent Finally the same strategy as was used at \roots\ = 7 TeV is
exploited to select inclusive $t$-channel events in a dataset  of
\intlum\ = 5.8~\invfb\ collected at \roots\ = 8
TeV~\cite{ref:tchan8TeV} and to estimate the corresponding backgrounds.
The corresponding $\sigma_{t,inc}$ value and the background normalizations are
derived with a binned maximum likelihood fit to the distribution of an
NN output (shown in fig.~\ref{fig:tchanXsec8TeV}) in two- and
three-jet samples. The NN exploits 11
kinematic variables to measure 
$\sigma_{t,inc} =  53.2 \pm 1.7\,\mathrm{(stat.)}\pm 10.6\,\mathrm{(syst.)\,pb}$.
With a relative uncertainty of 19\%, the result is still dominated by
radiation and jet energy scale (JES) systematic uncertainties.
 The value of
$|V_{tb}|$  is found to be $V_{tb} =  1.04^{+0.10}_{-0.11} $ and if
$|V_{tb}|$ is required  to be smaller than unity,  it is found that
$|V_{tb}| > 0.80$ at 95\% confidence level.
\vspace{-0.1cm}

\subsection{Inclusive single top quark Wt-channel cross section at \roots\  = 8 TeV}
\begin{figure}[h]
\centerline{
\includegraphics[width=0.43\textwidth]{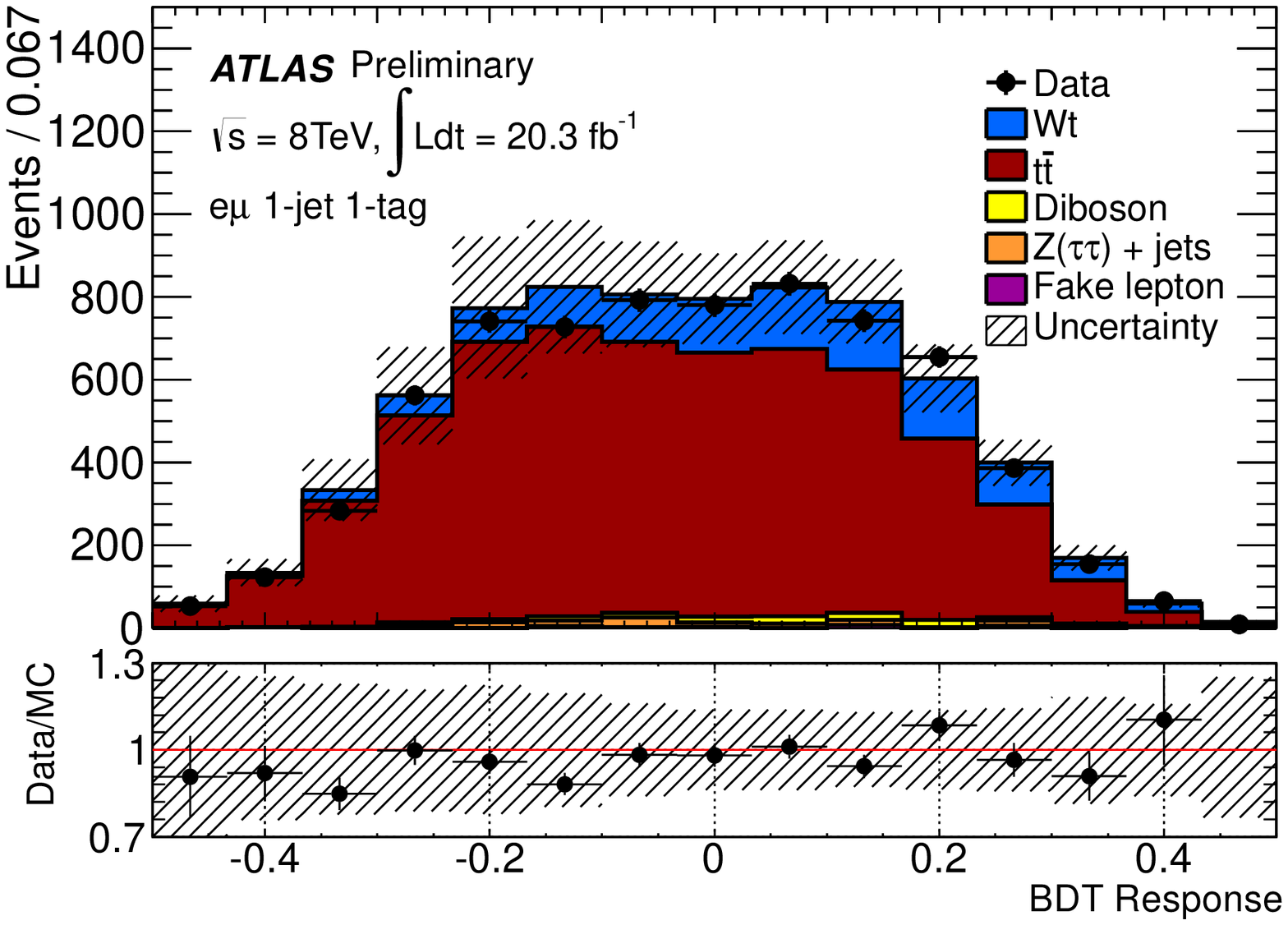}     
\includegraphics[width=0.43\textwidth]{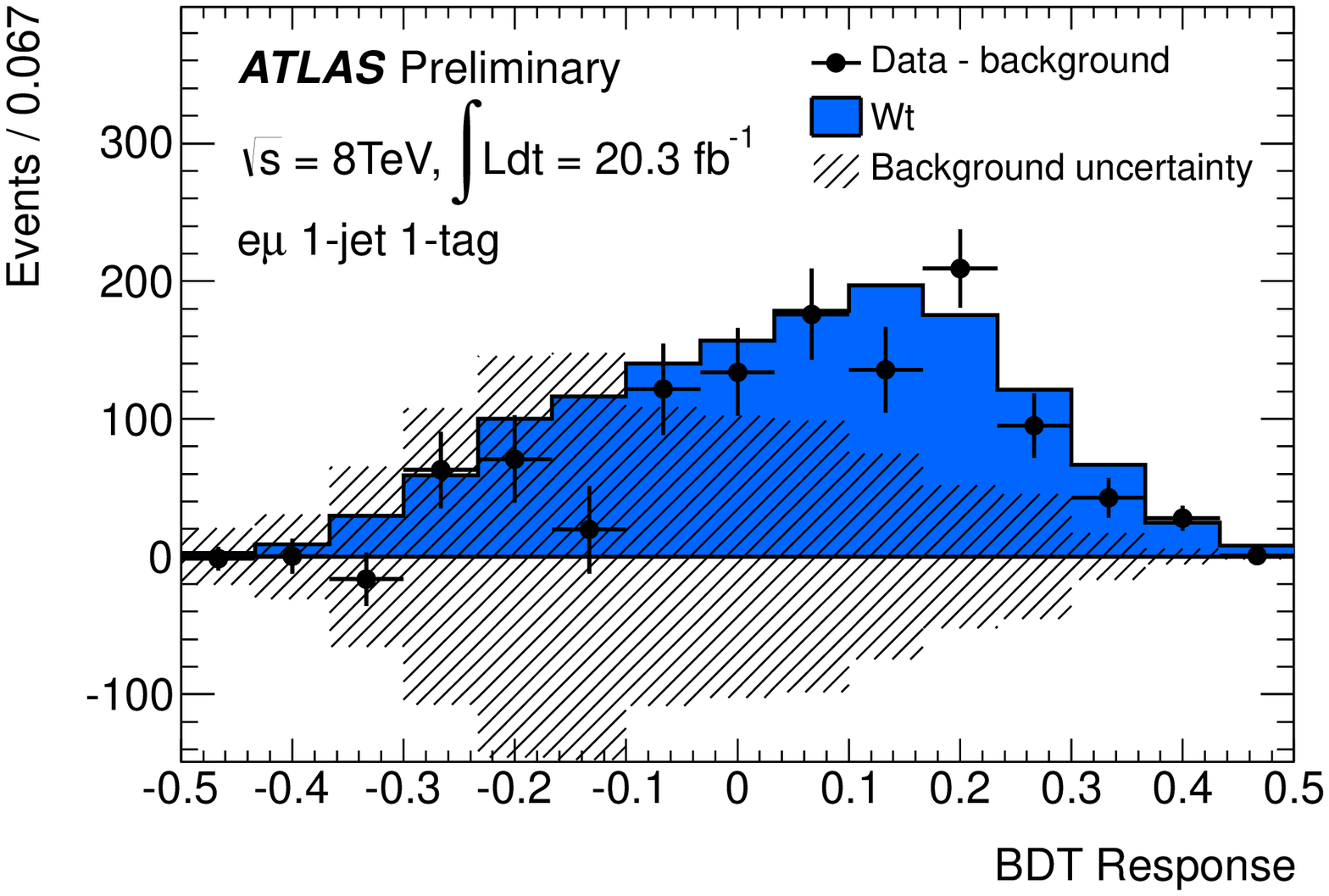}     
}
\caption{Distributions of the BDT classifier for 1-jet
  events before (left) and after (right) background subtraction after
  $Wt$ selection at \roots\ = 8 TeV~\cite{ref:WtChan8TeV}. Signal
  and backgrounds are normalized to the fit results. The hashed area is the quadratic sum of all background
  uncertainties.}
\label{fig:wt}
\end{figure}
The full dataset at \roots\ = 8 TeV is used to recognize $Wt$-single
top quark production~\cite{ref:WtChan8TeV}. Events are selected by requiring
an opposite sign $e$-$\mu$ pair with one or two high \pt\ central jets
and at least one $b$-tagged jet.  The dominant \ttbar\ background is
estimated from simulation like the small diboson and \zjets\
contributions, while fake leptons are derived with a matrix-method
data-driven technique. The $Wt$ cross section, $\sigma_{Wt}$  and the
background normalizations are derived using a binned maximum
likelihood fit to the distributions of the outputs of Boosted Decision
Trees (BDT) trained with 19 and 20 kinematic variables in samples with
one and two-jets, respectively. Fig.~\ref{fig:wt} shows the
data-prediction  comparison for the BDT output distribution (left) and
the background-subtracted result (right). The latter shows a clear
enhancement due to $Wt$ events standing out above the overall
background uncertainty. 
The resulting
$\sigma_{Wt} =  27.2 \pm 2.8\,\mathrm{(stat.)} \pm
5.4\,\mathrm{(syst.)\,pb}$ 
 provides a 4.2 standard deviation evidence for
$Wt$ single top quark production. The relative uncertainty of 22\%  is
dominated  by systematic effects relative to the generator and hadronization models.
The uncertainties related to $b$-tagging and two JES components are
constrained by the data with a profile-likelihood approach.
\setlength{\intextsep}{3pt}
\setlength{\columnsep}{6pt}
 \begin{wrapfigure}[18]{l}{0.42\textwidth}
\includegraphics[width=0.42\textwidth]{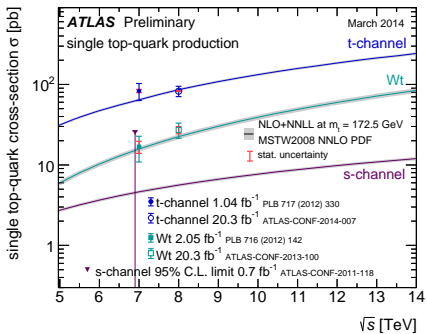}   
\caption{Summary of measurements of single top quark production
  cross-sections~\cite{ref:xsecCom8TeV} in various channels as a
  function of \roots\ compared  to calculations at the NLO QCD level
  complemented with NNLL resummation.  For the s-channel only an upper
  limit is shown.}
\label{fig:singleTopXsecSummary}
\end{wrapfigure}
The value of the CKM matrix element $|V_{tb}|$ is found to be $V_{tb}
=  1.10 \pm 0.12$, clearly consistent with the \roots\ = 7 and 8 TeV
results reported in sect.~\ref{sect:singleTop7TeV}.
If $|V_{tb}|$ is required  to be smaller than unity,  it is found
that $|V_{tb}| > 0.72$ at 95\% confidence level.

\vspace{-0.5cm}
\subsection{Single top production cross section summary}
ATLAS has been measuring single top quark production cross sections at both
\roots\ = 7 TeV and \roots\ = 8 TeV.
As summarized in fig.~\ref{fig:singleTopXsecSummary}, all the results are
consistent with SM predictions calculated at approximate
NNLO~\cite{ref:xsecCom8TeV}.
The $t$-channel production is observed and $\sigma_{t,inc}$ is
measured with a relative uncertainty ranging between 19\% and 25\%. 
There is evidence for $Wt$-channel production
with  $\sigma_{Wt}$  measured with 20\% to 30\%  relative uncertainty.
An upper limit at 95\% confidence level is placed on $s$-channel
production cross section using 0.7~\invfb\ of the full
4.7~\invfb\ at \roots\ = 7 TeV. Using about 5.8 \invfb\ of integrated luminosity collected at \roots\ = 8 TeV ATLAS also contributes to the first LHC
combination to derive
$\sigma_{t,inc}= 85 \pm 4\,\mathrm{(stat.)} \pm 11\,\mathrm{(syst.)} \pm 3\,\mathrm{(lumi)\,pb}$~\cite{ref:singleTopCom}
with an improved relative precision of 14\%.
\vspace{-0.4cm}
\section{Conclusions}
\vspace{-0.3cm}
The study of top quark production in ATLAS is in full swing
thanks to the combined high quality performance of both the detector and the LHC:
a very rich programme is well under way.
By exploiting the large dataset delivered by the LHC, ATLAS is
testing top quark strong and electroweak inclusive production at an
unprecedented precision level.
Consistency with the SM predictions is observed for all measurements of top quark strong and electroweak
production that use data collected by ATLAS in LHC proton-proton collisions 
at \roots\ = 7 and 8 TeV corresponding to integrated luminosities of up to about 4.7 \invfb\ and 20 \invfb,  respectively.
The \sigmattbar\ relative
uncertainty is as low as 4.8\%
compared to $\approx$ 4\%  relative uncertainty for the
latest NNLO+NNLL  predictions. The relative
uncertainties for $\sigma_{t,inc}$ and $\sigma_{Wt}$ are 19\%  and
25\% respectively.
There is still room for improvement for the observation of $s$-channel
single top quark events, so far only
searched for by using a fraction of the data available
at \roots\ = 7 TeV.
Differential cross section measurements provide major SM tests
in a completely new phase space  and complement new physics searches
with 10\%-20\% relative uncertainties, presently only using the full
dataset at \roots\ = 7 TeV.  
Updated results are expected for both inclusive and differential
cross sections using the full datasets at \roots\ = 7 and 8
TeV.
The uncharted kinematic phase space to be explored in LHC Run 2, 
starting in spring 2015,  will allow further improvement of these results by
exploiting the nearly tripled top quark production cross sections to
perform higher precision inclusive, exclusive and differential cross
section measurements.

\end{document}